\documentclass[preprint,amsmath,amssymb,showpacs,longbibliography]{revtex4-1}
\usepackage{hyperref}
\pdfoutput=1
\usepackage{graphicx}

\usepackage{amsmath}
\usepackage{bm}
\usepackage{amssymb}
\usepackage{textcomp}
\usepackage{stmaryrd} 
\usepackage{ulem}

\usepackage{import}
\usepackage{color}
\usepackage{verbatim}

\usepackage{float}

\makeatletter
\DeclareRobustCommand{\element}[1]{\@element#1\@nil}
\def\@element#1#2\@nil{%
  #1%
  \if\relax#2\relax\else\MakeLowercase{#2}\fi}
\makeatother

\def\XXint#1#2#3{{\setbox0=\hbox{$#1{#2#3}{\int}$}
     \vcenter{\hbox{$#2#3$}}\kern-.5\wd0}}

\begin{document}

\title{Splitting of Interlayer Shear Modes and Photon Energy Dependent Anisotropic Raman Response in $N$-Layer ReSe$_2$ and ReS$_2$}

\author{Etienne Lorchat}
\affiliation{Institut de Physique et Chimie des Mat\'eriaux de Strasbourg and NIE, UMR 7504, Universit\'e de Strasbourg and CNRS, 23 rue du L\oe{}ss, BP43, 67034 Strasbourg Cedex 2, France}

\author{Guillaume Froehlicher}
\affiliation{Institut de Physique et Chimie des Mat\'eriaux de Strasbourg and NIE, UMR 7504, Universit\'e de Strasbourg and CNRS, 23 rue du L\oe{}ss, BP43, 67034 Strasbourg Cedex 2, France}

\author{St\'ephane Berciaud}
\email{stephane.berciaud@ipcms.unistra.fr}
\affiliation{Institut de Physique et Chimie des Mat\'eriaux de Strasbourg and NIE, UMR 7504, Universit\'e de Strasbourg and CNRS, 23 rue du L\oe{}ss, BP43, 67034 Strasbourg Cedex 2, France}


\begin{abstract}
We investigate the interlayer phonon modes in $N$-layer rhenium diselenide (ReSe$_2$) and rhenium disulfide (ReS$_2$) by means of ultralow-frequency micro-Raman spectroscopy. These transition metal dichalcogenides exhibit a stable distorted octahedral (1T') phase with significant in-plane anisotropy, leading to sizable splitting of  the (in-plane) layer shear modes. The fan-diagrams associated with the measured frequencies of the interlayer shear modes and the (out-of-plane) interlayer breathing modes are perfectly described by a finite linear chain model and allow the determination of the interlayer force constants. Nearly identical values are found for ReSe$_2$ and ReS$_2$. The latter are appreciably smaller than but on the same order of magnitude as the interlayer force constants reported in graphite and in trigonal prismatic (2Hc) transition metal dichalcogenides (such as MoS$_2$, MoSe$_2$, MoTe$_2$, WS$_2$, WSe$_2$), demonstrating the importance of van der Waals interactions in $N$-layer ReSe$_2$ and ReS$_2$. In-plane anisotropy results in a complex angular dependence of the intensity of all Raman modes, which can be empirically utilized to determine the crystal orientation. However, we also demonstrate that the angular dependence of the Raman response drastically depends on the incoming photon energy, shedding light on the importance of resonant exciton-phonon coupling in ReSe$_2$ and ReS$_2$.




\end{abstract}
%

\maketitle


Although well-documented in their bulk forms for decades, transition metal dichalcogenides (TMD)~\cite{Wilson1969} have recently attracted considerable attention since the observation of intense photoluminescence (PL) from monolayers of molybdenum disulfide (MoS$_2$)~\cite{Mak2010,Splendiani2010}. So far, the majority of works on TMDs has focused on trigonal prismatic compounds, with AbA/BaB stacking (2Hc phase~\cite{Ribeiro2014})  such as MoS$_2$, MoSe$_2$, MoTe$_2$, WS$_2$, WSe$_2$~\cite{Mak2010,Splendiani2010,Tonndorf2013,Zhao2012,Ruppert2014,Lezama2015}. The latter exhibit an indirect-to-direct bandgap transition when their thickness is reduced down to the monolayer limit and hold promise for opto-electronic~\cite{Wang2012} and valleytronic applications~\cite{Xu2014}. In contrast to these high-symmetry structures, other semiconducting TMD, such as rhenium disulfide (ReS$_2$)~\cite{Tongay2014,Corbet2015,Zhang2015,Feng2015,Liu2015,Lin2015} and rhenium diselenide (ReSe$_2$)~\cite{Yang2014,Wolverson2014,Zhao2015} (thereafter denoted ReX$_2$)  exhibit a distorted octahedral phase (denoted 1T') with much lower symmetry and significant in-plane anisotropy~\cite{Wilson1969,Ho1997,Ho1998,Ho2004,Tiong1999}. $N$-layer ReX$_2$ have been shown to remain indirect bandgap semiconductors, with faint PL irrespective of the number of layers $N$~\cite{Tongay2014,Zhao2015}. In spite of these poor emission properties, ReX$_2$ offer exciting perspectives for anisotropic electronic devices~\cite{Tiong1999,Liu2015,Lin2015}. However, such developments require unambiguous fingerprints of in-plane anisotropy as well as fast and accurate methods to determine the crystal orientation, the strength of interlayer interactions, and the stacking order. 

Recently, Raman scattering studies of ReX$_2$  have revealed that the Raman response of these anisotropic systems depends subtly on the relative orientation of the crystal axis and of the exciting laser field and also on the angle between the polarizations of the incoming and Raman scattered photons~\cite{Wolverson2014,Feng2015,Chenet2015,Zhao2015,Nagler2015}. Nevertheless, most studies on ReX$_2$ have focused on the complex manifold of intralayer Raman modes, which exhibits minor variations as a function of the number of layers~\cite{Wolverson2014,Feng2015,Chenet2015}.  In contrast, the low-frequency interlayer shear modes (LSM) and breathing modes (LBM) in TMD make it possible to unambiguously determine the number of layers and evaluate the strength of van der Waals interlayer coupling~\cite{Zhao2015,Nagler2015,Plechinger2012,Zeng2012,Zhao2013,Zhang2013,Boukhicha2013,Froehlicher2015}. In the case of $N$-layer ReX$_2$, a splitting of the (in-plane) LSM is expected~\cite{Zhao2015} and should provide an invaluable tool to investigate in-plane anisotropy. However an experimental observation of this splitting is still lacking.

\begin{figure*}[!tbh]
\begin{center}
\includegraphics[width=1\linewidth]{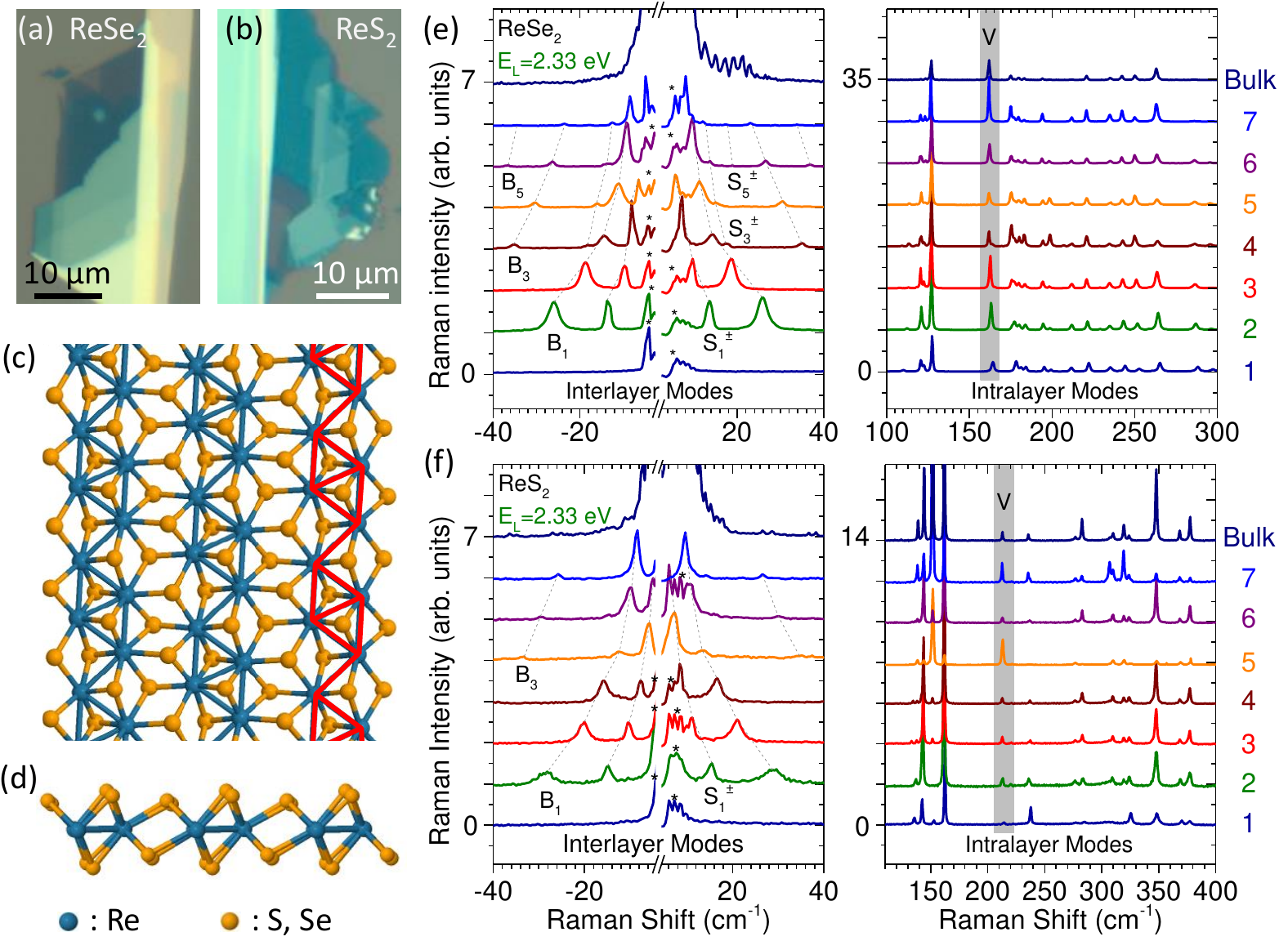}
\caption{Optical image of a $N$-layer (a) ReSe$_2$ and (b) ReS$_2$ crystal. (c) Top view and (d) side view of a 1T'-ReX$_2$ monolayer (the rhenium chains are highlighted in red). Raman spectra recorded on (e) $N$-layer ReSe$_2$ and (f) $N$-layer ReS$_2$ at $E_{\rm L}=2.33~\rm eV$ in the parallel polarization (XX) configuration. The dashed lines track the frequencies of layer shear modes (LSM) and layer breathing modes (LBM). Asterisks mark residual contributions from the exciting laser beam and the fifth intralayer mode (V) is highlighted with a grey rectangle.)}
\label{Fig1}
\end{center}
\end{figure*}

In this article, we report a detailed Raman study of $N$-layer ReSe$_2$ and ReS$_2$. We resolve a clear splitting of the (in-plane) LSM. In contrast the (out-of plane) LBM do not exhibit any splitting, allowing an unambiguous distinction between LSM and LBM in the Raman spectra. We report the fan diagrams associated with all the Raman active LSM and LBM up to $N=7$. The observed branches of LSM and LBM are very well described using a simple linear chain model. Noteworthy, in contrast to the case of 2Hc-TMD, the frequency of both LBM \textit{and} LSM seemingly decrease with increasing $N$ and we rationalize this observation using a symmetry argument. Finally, angle-dependent Raman studies of the LSM and LBM in bilayer ReX$_2$ recorded at two different laser energies reveal complex patterns which, on the one hand, can empirically be used to identify the crystal orientation but, on the other hand, shed light on the importance of resonance Raman effects in these strongly anisotropic layered materials.

\section*{Results and discussion}

\begin{figure*}[!htb]
\begin{center}
\includegraphics[width=0.8\linewidth]{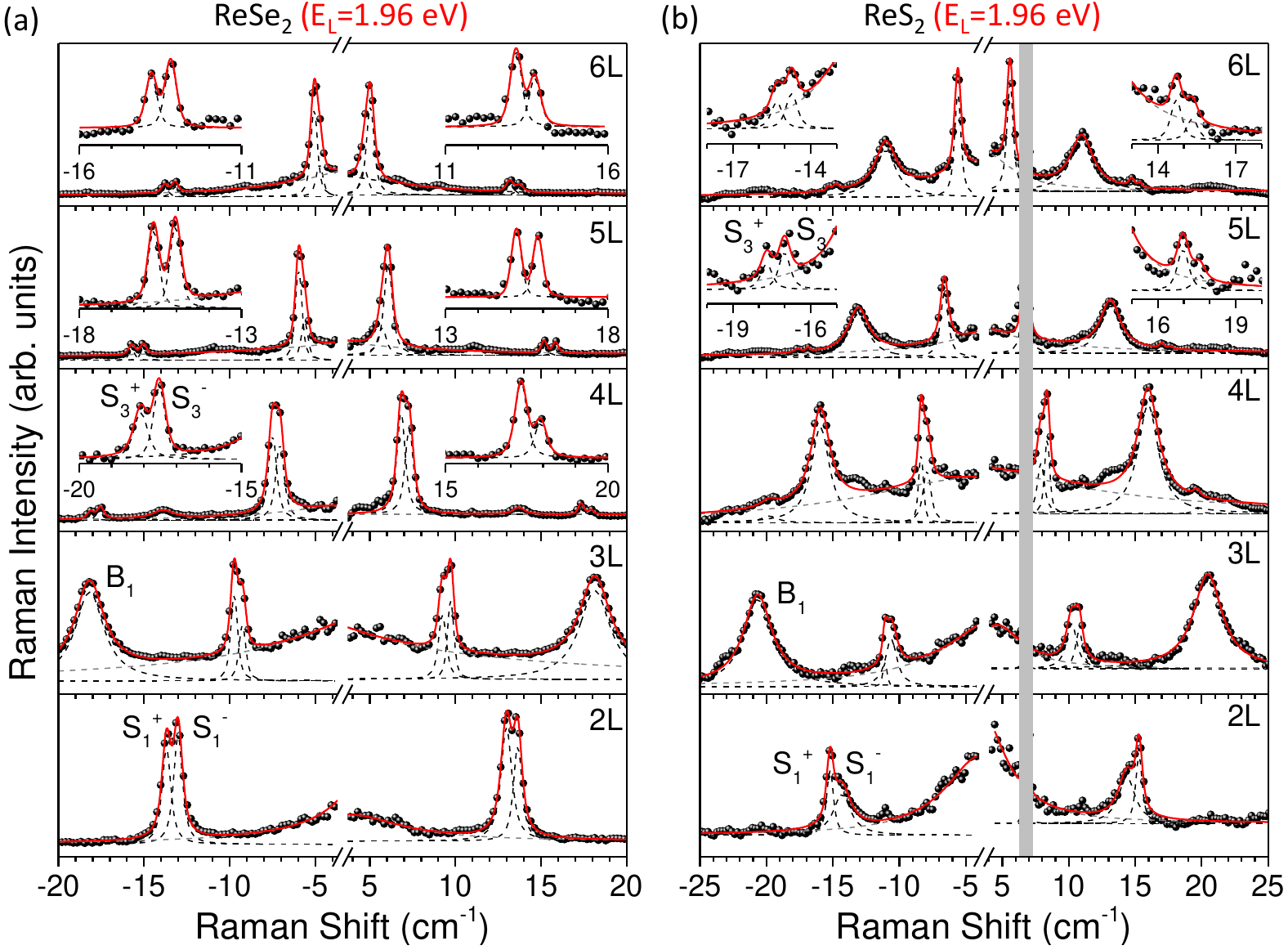}
\caption{Ultralow frequency Raman spectra of $N=2$ to $N=6$ layer (a) ReSe$2$ and (b) ReS$_2$, recorded at $E_{\rm L}=1.96~\rm eV$ in the parallel (XX) configuration. The experimental data (black spheres) are fit to Voigt profiles (red lines). The grey bar in (b) masks a residual contribution from the laser beam. The mode labels correspond to the notations in Fig.~\ref{Fig3}.}
\label{Fig2}
\end{center}
\end{figure*}

\paragraph*{\textbf{\textit{Interlayer phonon modes in ReSe$_2$ and ReS$_2$}}}


Figure~\ref{Fig1}(a),(b) shows optical images of $N$-layer ReSe$_2$ and ReS$_2$, respectively. The 1T' structure of monolayer  ReX$_2$ is shown in Figure~\ref{Fig1}(c),(d). In contrast to the 1T phase, the 1T' phase displays covalent bonding between metal (here Re) atoms. The covalently bound Re atoms form diamond-like patterns that are connected to each other, leading to quasi one-dimensional Re chains~\cite{Wilson1969,Tongay2014,Feng2015,Zhao2015}. In practice, ReX$_2$ crystals preferentially cleave along these chains (see Figure~\ref{Fig1}(a),(b)). The 1T' phase has considerably lower symmetry than the 1H phase~\cite{Feng2015,Tongay2014,Wilson1969}. Monolayer 1T'-ReX$_2$ belongs to the $C_i$ space group and is only endowed with one inversion center. The monolayer unit cell of ReX$_2$ has 12 atoms giving rise to 36 zone-center phonon modes. These modes have $A_u$ or $A_g$ symmetry and are infrared or Raman active, respectively~\cite{Feng2015}. In monolayer ReX$_2$, one thus expects 18 Raman-active modes.

Figure~\ref{Fig1}(e),(f) shows the Raman spectra of $N=1$ to $N=7$ layer and bulk ReSe$_2$ and ReS$_2$, respectively, recorded at $E_{\rm L}=2.33~\rm eV$ in the parallel (XX) configuration for randomly oriented samples. As previously reported~\cite{Wolverson2014,Tongay2014,Feng2015}, we can identify the complex manifold of intralayer modes in the range $100-400~\rm{cm^{-1}}$.  In the following, we will only consider the fifth lowest frequency intralayer mode (denoted mode V as in Ref.~\citenum{Chenet2015})  at approximately $162~\rm cm^{-1}$ (respectively $213~\rm cm^{-1}$) in ReSe$_2$ (respectively ReS$_2$). This mode predominantly involves atomic displacement of the Re-Re bonds~\cite{Feng2015}.

We now concentrate on the low-frequency Raman response of $N$-layer ReX$_2$, at Raman shifts below 40~cm$^{-1}$ (see left panels in Fig.~\ref{Fig1}(e),(f) and Fig.~\ref{Fig2}). For increasing $N\geq2$, one observes well-defined stokes and anti-stokes peaks that we assign to the LSM and LBM branches.  As expected, no interlayer mode-features are observed for $N=1$. Only two interlayer mode-features are resolved for $N=2$ and $N=3$. As in 2Hc-TMD, two new features appear at $N=4$ (see also Figure~\ref{Fig2}) and again at $N=6$. Interestingly, the frequency of all the observed LSM and LBM seemingly decreases when $N$ increases, and ultimately merge into the Rayleigh background for large $N$. This observation is in stark contrast with 2Hc-TMD, where the frequencies of the LSM (respectively, LBM) have been shown to increase (respectively, decrease) with increasing $N$~\cite{Zhao2013,Zhang2013,Froehlicher2015} and complicates the identification of the LSM and LBM. Remarkably, bulk ReX$_2$ crystals do not display any signature of finite frequency interlayer modes, as expected from group theory predictions for the 1T' structure~\cite{Ribeiro2014,Feng2015}. Remarkably, bulk ReX$_2$ crystals do not display any signature of finite frequency interlayer modes. This observation is consistent with a bulk 1T' structure with one ReX$_2$ layer per bulk unit cell~\cite{Ribeiro2014,Feng2015} and contrasts with the case of bulk 2Hc-TMD, which have two layers per bulk unit cell and display a Raman active LSM~\cite{Ribeiro2014}.

In the spectra shown in Figure~\ref{Fig1}(e),(f) no clear splitting of the interlayer mode-features can be resolved. Figure~\ref{Fig2} shows higher resolution spectra recorded at a lower photon energy of $E_{\rm L}=1.96~\rm eV$. In both ReS$_2$ and ReSe$_2$, peaks belonging to the lowest- and third-lowest frequency branches display small but measurable splittings of $\lesssim 1~\rm cm^{-1}$, while the features belonging to second-lowest frequency branch do not split. Based on these observations, we assign the branches that do not exhibit any splitting to LBM and the branches that split to LSM, respectively. We note that our assignment is consistent with previous measurements and calculations on layered compounds, which, for a given normal mode, systematically reported lower frequencies for LSM than for the LBM~\cite{Michel2012,Zhao2013,Boukhicha2013,Zhang2013,Froehlicher2015}. In addition, our direct measurement of the LSM splittings in ReSe$_2$ are in-line with recent density functional theory (DFT) calculations~\cite{Zhao2015}.

\paragraph*{\textbf{\textit{Fan diagrams and interlayer force constants}}}

\begin{figure}[!htb]
\begin{center}
\includegraphics[width=0.75\linewidth]{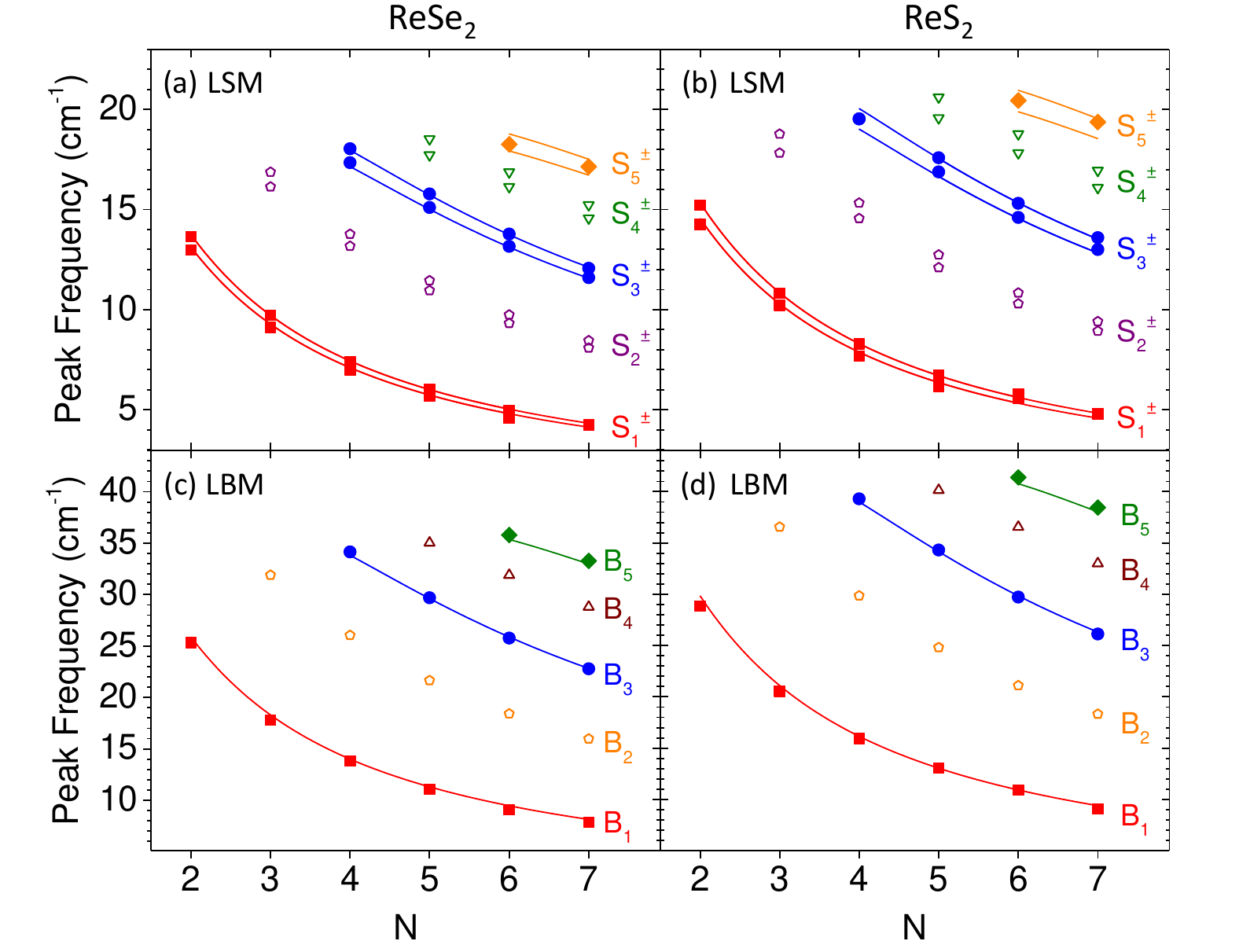}
\caption{Frequencies of the Raman-active interlayer shear modes ($S_1^{\pm}$, $S_3^{\pm}$,\dots) of (a) ReSe$_2$ and (b) ReS$_2$ and of the interlayer breathing modes ($B_1^{~}$, $B_3^{~}$,\dots) of (c) ReSe$_2$ and (d) ReS$_2$ as a function of the number of layers $N$ (solid symbols). The measured frequencies are fit to Eq.~\ref{eqLCM} (solid lines). Open symbols represent the infrared-active modes ($S_2^{\pm}$, $S_4^{\pm}$,\dots, and $B_2^{~}$, $B_4^{~}$,\dots) also predicted by Eq.~\ref{eqLCM}.}
\label{Fig3}
\end{center}
\end{figure}

Based on these observations, we may now plot, in Fig.~\ref{Fig3} the fan diagrams associated with the two non-degenerate LSM and with the LBM in $N$-layer ReX$_2$.  As previously introduced for other layered materials~\cite{Tan2012,Zhao2013,Zhang2013,Boukhicha2013,Froehlicher2015,Nagler2015}, a linear chain model can be used to fit the frequencies of the observed modes and extract the interlayer force constants. The LSM and LBM frequencies write
\begin{equation}
 \omega_{S_k^{\pm},B_k}^{~}\left(N\right)=\widetilde{\omega}_{S^{\pm}_{~},B}^{~}\;\sqrt{1-\cos{\left(\frac{k\pi}{N}\right)}}
\label{eqLCM}
\end{equation}
with $k\in\llbracket 1, N-1\rrbracket $ ($k=0$ corresponds to the zero frequency acoustic mode at $\Gamma$). The observed Raman-active modes correspond to branches  with $k=1,\;3,\;5,\dots$ for the LSM and LBM. The latter are denoted $S_1^{\pm}$, $S_3^{\pm}$, $S_5^{\pm}$,\dots and $B_1^{}$, $B_3^{}$, $B_5^{}$,\dots, respectively (see Fig.~\ref{Fig3}). In Eq.~\ref{eqLCM}, $\widetilde{\omega}_{S^{\pm}_{~},B}^{~}=\sqrt{\frac{2\:\kappa_{S^{\pm},B}}{\mu}}$ denotes the non-degenerate LSM ($S_1^{\pm}$) frequencies and the LBM ($B_1$) frequency for bilayer ReX$_2$, with $\kappa_{S^{\pm},B}^{~}$ the associated interlayer force constant and $\mu$ the mass per unit area in ReX$_2$. The resulting force constants are reported in Table~\ref{TabFC} and compared to measurements on 2Hc-TMD.

\begin{figure}[!htb]
\begin{center}
\includegraphics[width=0.6\linewidth]{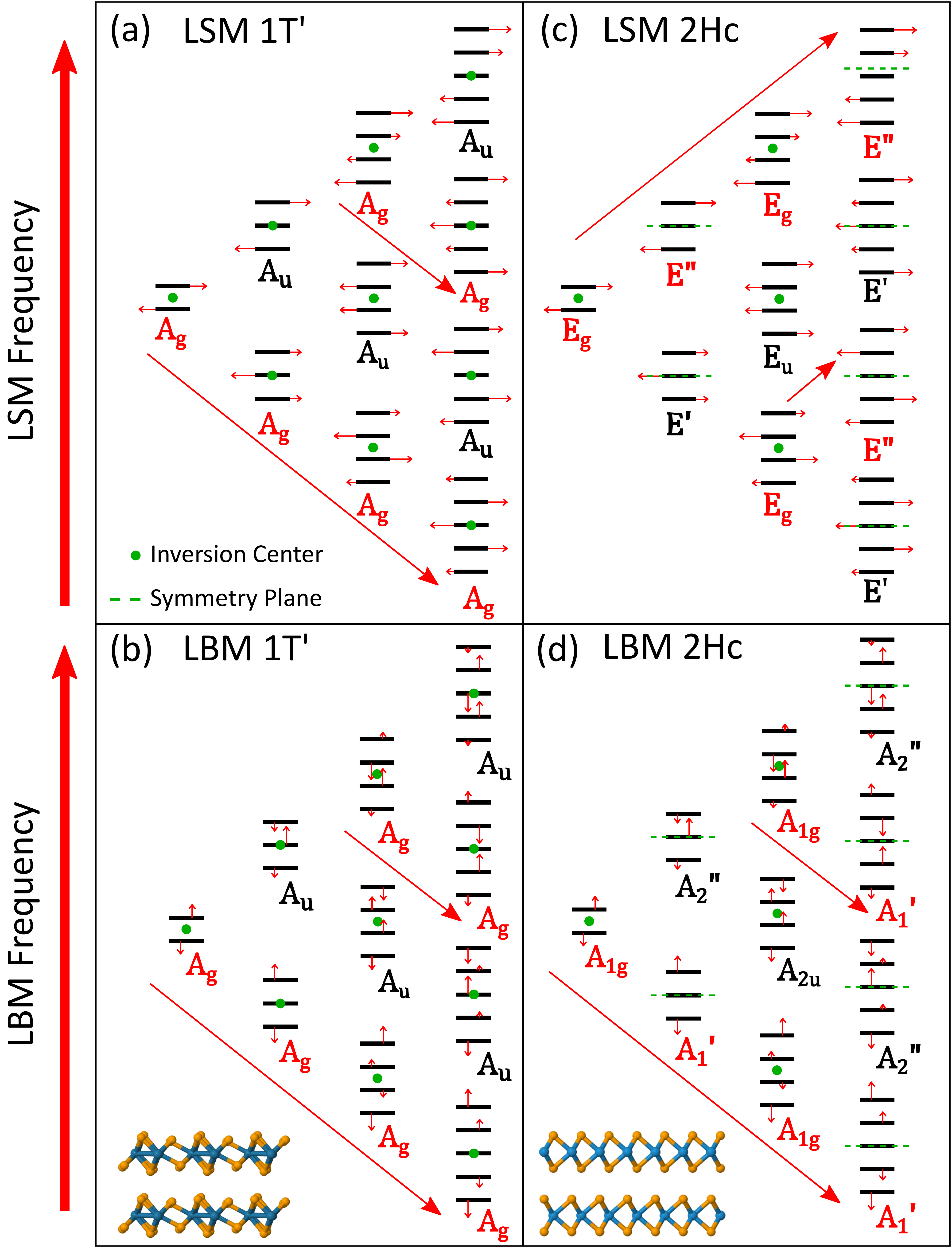}
\caption{Normal displacements associated with the interlayer shear and breathing modes calculated using the finite linear chain model for layered materials exhibiting a 1T' (a,b) or 2Hc (c,d) phase (see side views of bilayer 1T'-ReSe$_2$ and 2Hc-MoS$_2$ in (c) and (d), respectively). The presence of an inversion center (for any value of $N$ in the 1T' phase and for even $N$ in the 2Hc phase) or of a mirror symmetry plane (for odd $N$  in the 2Hc phase) is indicated. The irreducible representation of each mode is indicated, with Raman-active modes in red, infrared active modes and Raman modes that are not observable in a backscattering geometry in black. The oblique arrows symbolize the decrease (increase) of the frequency of one given mode as $N$ increases.}
\label{Fig4}
\end{center}
\end{figure}

\setlength{\tabcolsep}{0.1cm}
\renewcommand{\arraystretch}{2}
\begin{table} [!htb]
\begin{center}
\begin{tabular}{ccc}

\hline
\hline

Material & $\kappa_{S^{\pm}}^{~} (10^{18}~\rm N/m^3)$ & $\kappa_{B}^{~} (10^{18}~\rm N/m^3)$   \\

\hline

ReSe$_2$ (this work) & 17.8/19.4 & 69   \\

ReS$_2$ (this work) & 17.1/18.9 & 69    \\

MoS$_2$ (Refs.~\citenum{Zhang2013,Boukhicha2013,Zhao2013}) & 28 & 87    \\

MoTe$_2$ (Ref.~\citenum{Froehlicher2015}) & 34 & 77   \\

WSe$_2$ (Ref.~\citenum{Zhao2013}) & 31 & 86   \\

Graphite (Ref.~\citenum{Tan2012,Lui2014}) & 13 & 94  \\

\hline

\end{tabular}
\end{center}  

\caption{Force constants per unit area of the low-frequency (LSM, LBM) modes in 1T' and 2Hc transition metal dichalcogenides, and graphite, extracted from a fit to a finite linear chain model (see Eq.~\eqref{eqLCM}).}
\label{TabFC}
\end{table}

The interlayer force constants in 1T'-ReX$_2$ appear to be significantly smaller, yet on the same order of magnitude as in 2Hc-TMD. One may even notice that $\kappa_{S^{\pm}}^{~}$ is larger in ReX$_2$ than in graphite~\cite{Tan2012}.  As a result, albeit the modest variations of the high-frequency intralayer Raman response\cite{Wolverson2014} and of the PL spectra~\cite{Tongay2014,Zhao2015}  on $N$ may suggest marginal interlayer coupling in ReX$_2$, our direct investigation of interlayer Raman modes establishes that $N$-layer ReX$_2$ must be regarded as a van der Waals coupled system, with obvious fingerprints of interlayer interactions.

We now propose a simple symmetry analysis to explain why the frequency of the LSM and LBM from a given branch both decrease as $N$ augments, in contrast to the case of 2Hc-TMD~\cite{Zhao2013,Zhang2013,Froehlicher2015}.  First, in keeping with recent studies in twisted TMD bilayers and van der Waals heterostructures made of 2Hc-TMD, we would like to point out that the observation of prominent LSM features in Fig.~\ref{Fig1} and~\ref{Fig2} suggests that there is a well-defined stacking order in our $N$-Layer ReSe$_2$ and ReS$_2$ samples~\cite{Lui2015,Puretzky2016}. Let us then assume that $N$-layer ReSe$_2$ and ReS$_2$ both have a 1T' structure, which has no mirror symmetry plane but preserves an inversion center for any value of $N$. In contrast, $N$-layer 2Hc compounds exhibit a mirror symmetry plane (but no inversion center) for odd $N$ and an inversion center (but no mirror symmetry plane) for even $N$~\cite{Zhao2013,Ribeiro2014}. Then, as illustrated in Figure~\ref{Fig4}, the (symmetric) $A_g$ Raman-active LSM and LBM in ReX$_2$ correspond the lowest frequency mode, third-lowest frequency mode,\dots, leading to branches that soften as $N$ increases from $\widetilde{\omega}_{S^{\pm}_{~},B}^{~}$ for $N=2$ down to zero frequency. In contrast, for 2Hc compounds, the Raman-active LSM correspond to the highest frequency mode, third-highest frequency mode,\dots, leading to branches that stiffen with increasing $N$, from $\widetilde{\omega}_{S^{\pm}_{~},B}^{~}$ for $N=2$ up to $\sqrt{2}\widetilde{\omega}_{S^{\pm}_{~},B}^{~}$ in the bulk limit, whereas the Raman-active LBM correspond to the lowest, third lowest,\dots frequency mode, leading to similar trends as for ReX$_2$.

\paragraph*{\textbf{\textit{Angular dependence of the Raman response}}}

\begin{figure*}[!hbt]
\begin{center}
\includegraphics[width=0.95\linewidth]{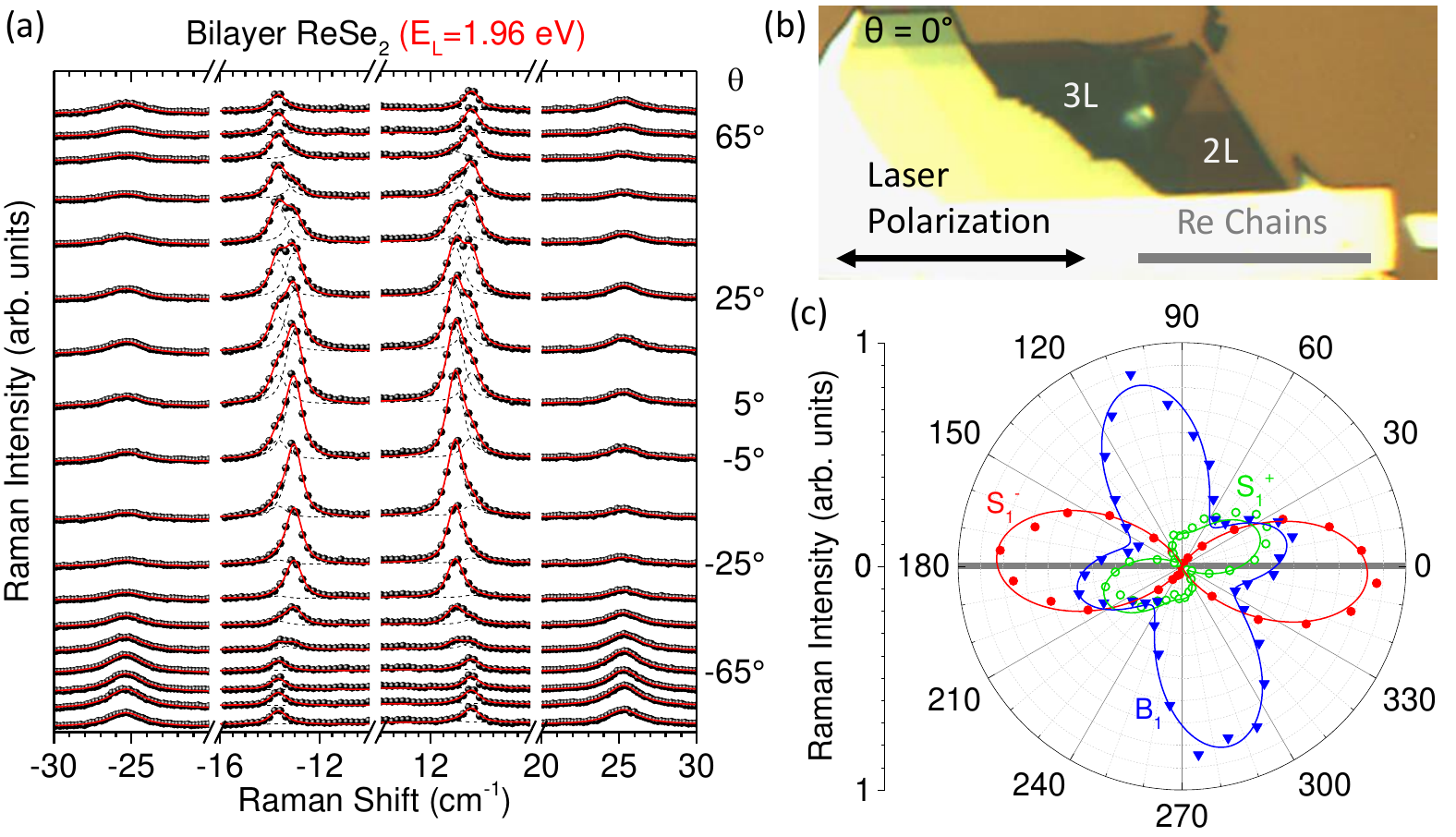}
\caption{(a) Angular dependence of the low-frequency Raman spectrum of a ReSe$_2$ bilayer recorded as a function of the angle $\theta$ between the linearly polarized laser field and the Re chains at $E_{\rm L}=1.96~\rm eV$ in the parallel (XX) configuration. The raw spectra (spheres) are fit to Voigt profiles (red lines) and are vertically offset for clarity. (b) Optical image of the ReSe$_2$ bilayer showing the direction of the rhenium chains. At $\theta=0^{\circ}$, the laser polarization is parallel to the Re chains. (c) Polar plots of the integrated Raman intensity of the $S_1^{\pm}$ ($S_1^{-}$ with red circles, $S_1^{+}$ with green open circles ) and $B_1^{~}$ modes (blue triangles). The solid lines in (c) are fits to the experimental data considering a Raman tensor with complex elements (see Supporting Information).}
\label{Fig5}
\end{center}
\end{figure*}

\begin{figure*}[!thb]
\begin{center}
\includegraphics[width=0.95\linewidth]{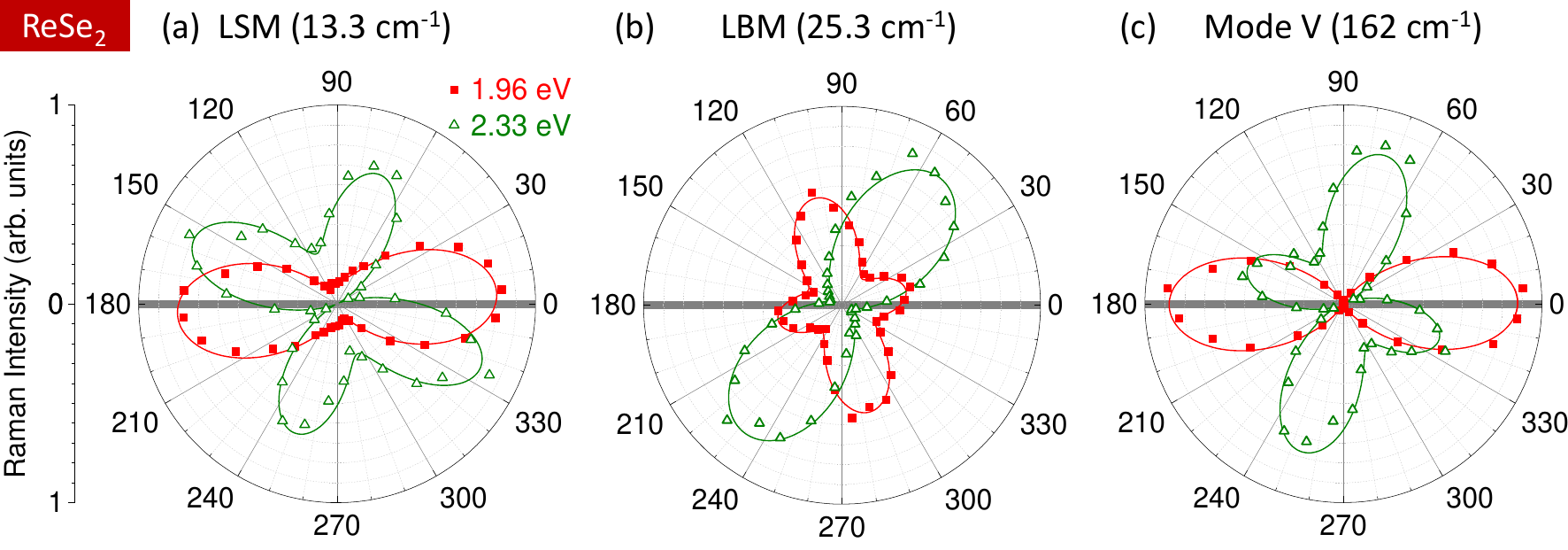}
\caption{Polar plots of the integrated Raman intensity of (a) the sum of the $S_1^{\pm}$ modes, (b) the $B_1^{~}$ mode, and (c) the mode V, recorded as a function of the angle $\theta$ between the linearly polarized laser field and the Re chains (solid grey lines in (a)-(c)), at two different photon energies $E_{\rm L}=1.96~\rm eV$ (red squares) and $E_{\rm L}=2.33~\rm eV$ (open green triangles) in the parallel (XX) configuration on the ReSe$_2$ bilayer shown in Fig.~\ref{Fig5}. The angular patterns are normalized for a clearer comparison. The solid lines are fits to the experimental data considering a Raman tensor with complex elements (see Supporting Information).}
\label{Fig6}
\end{center}
\end{figure*}

\begin{figure*}[!hbt]
\begin{center}
\includegraphics[width=0.95\linewidth]{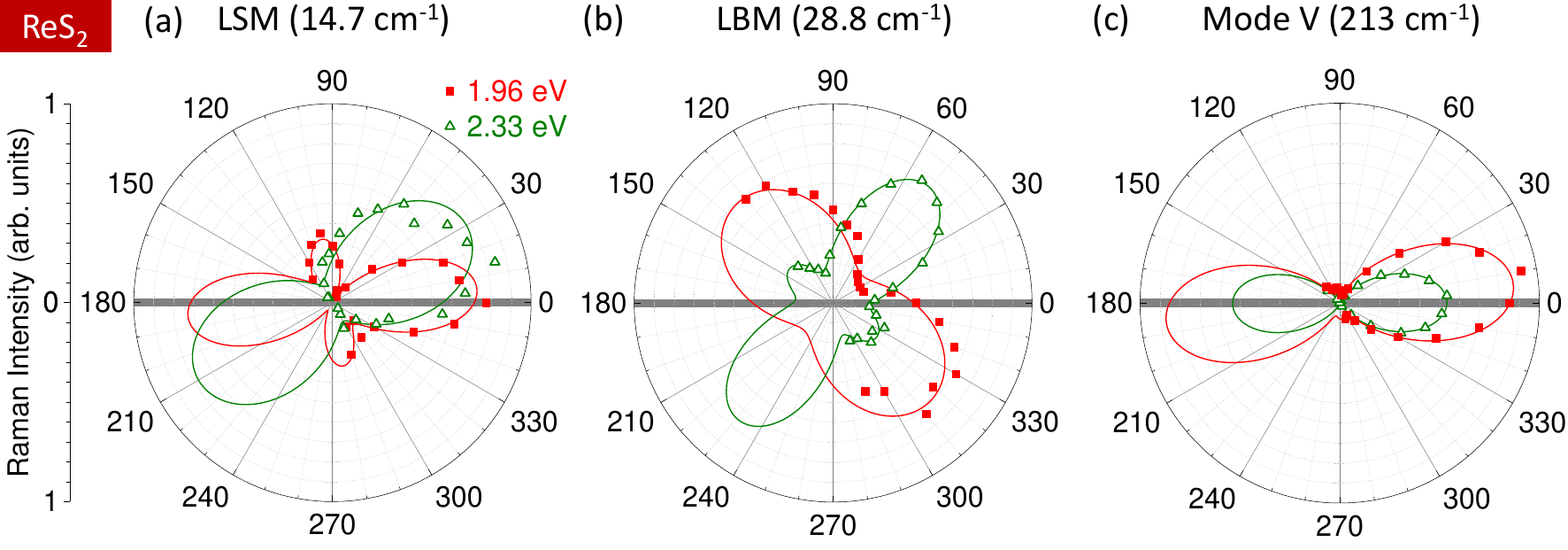}
\caption{Polar plots of the integrated Raman intensity of (a) the sum of the $S_1^{\pm}$ modes, (b) the $B_1^{~}$ mode, and (c) the mode V recorded as a function of the angle $\theta$ between the linearly polarized laser field and the Re chains (solid grey lines in (a)-(c)), at two different photon energies $E_{\rm L}=1.96~\rm eV$ (red squares) and $E_{\rm L}=2.33~\rm eV$ (open green triangles) in the parallel (XX) configuration on a same ReS$_2$ bilayer. The angular patterns are normalized for a clearer comparison.  The solid lines  are fits to the experimental data considering a Raman tensor with complex elements (see Supporting Information).}
\label{Fig7}
\end{center}
\end{figure*}

Next, we address the angular dependence of the Raman response in ReX$_2$. Figure~\ref{Fig5} displays the evolution of the LSM (\textit{i.e.}, the $S_1^{\pm}$ modes) and LBM (\textit{i.e.}, the $B_1^{~}$ mode) in bilayer ReSe$_2$, recorded as a function of the angle $\theta$ between the linearly polarized laser electric field and the Re chains at $E_{\rm L}=1.96~\rm eV$ in the XX configuration. Following the empirical method proposed by Chenet \textit{et al.} for ReS$_2$, the orientation of the Re chains in ReSe$_2$ has been preliminarily determined by inspecting the angular dependence of the integrated intensity of the mode V ($I_{\rm V}^{~}$) introduced above~\cite{Chenet2015} (see Fig.~\ref{Fig1}), allowing us to define the reference angle $\theta=0^{\circ}$ for a laser polarization parallel to the Re chains (see solid grey lines in Fig.~\ref{Fig5}-~\ref{Fig7} and Supporting Information). As $\theta$ varies, one clearly observes large, $\pi$-periodic variations of the LSM and LBM integrated intensities (denoted $I_{ S_1^{\pm}}$ and $I_{ B_1^{~}}$, respectively), which are direct fingerprints of in-plane anisotropy. The frequencies and linewidth of the $S_1^{\pm}$ and $B_1$-mode features are found to be independent on $\theta$.
 
Remarkably, as shown in Fig.~\ref{Fig5}(c), we notice that at $E_{\rm L}=1.96~\rm{eV}$, $I_{ S_1^-}$ reaches a maximum at $\theta\approx 0^{\circ} \pmod{180^{\circ}}$ and exhibits approximately a $\cos^4\theta$ dependence, whereas $I_{ S_1^+}$ displays a more complex angular dependence with a maximum at $\theta\approx 25^{\circ} \pmod{180^{\circ}}$ and secondary maxima near $\theta\approx 105^{\circ} \pmod{180^{\circ}}$, close enough to the direction perpendicular to the Re chains. Finally, the polar plot of $I_{ B_1^{~}}$ also displays prominent maxima at $\theta=105^{\circ} \pmod{180^{\circ}}$, as well as secondary transverse maxima at $\theta=25^{\circ} \pmod{180^{\circ}}$ that also match the maxima of $I_{ S_1^+}$.

Interestingly, as it has also been reported in few-layer ($> 5~\rm nm$-thick) black phosphorus films~\cite{Kim2015}, the observed angular patterns appear to be strongly dependent of the laser photon energy. Indeed, as shown in Fig.~\ref{Fig6}  and further outlined in the Supporting Information, the same measurements performed at a slightly higher photon energy of $E_{\rm L}=2.33~\rm eV$ on the same ReSe$_2$ bilayer reveal completely different angular dependencies for $I_{ S_1^{\pm}}$, $I_{ B_1^{~}}$, $I_{\rm V}$ and, importantly, no Raman intensity maximum at $\theta=0^{\circ}$ for any of these modes. Remarkably, we find that the total integrated intensity of the $S_1$  mode ($I_{S_1^{~}}=I_{ S_1^{-}}+I_{ S_1^+}$) and $I_{\rm V}$ display similar patterns at both photon energy. Remarkably, the modes  that exhibit \textit{two-lobe} angular patterns at $E_{\rm L}=1.96~\rm eV$ (\textit{i.e.}, the $S_1$ mode and mode V), exhibit \textit{four-lobe} patterns at $E_{\rm L}=2.33~\rm eV$ and \textit{vice versa} (cf. the $B_1^{~}$ mode).
Furthermore, as shown in Fig.~\ref{Fig7}, we performed a similar study on a ReS$_2$ bilayer with known crystal orientation  and found angular patterns for the Raman intensities that bear no particular similarity with their counterparts in ReSe$_2$. The angular patterns of $I_{S_1}$ and $I_{B_1}$ depend on the incoming photon energy, whereas similar angular patterns are observed for $I_{\rm V}$ at $E_{\rm L}=1.96~\rm eV$ and $E_{\rm L}=2.33~\rm eV$, with a maxima in the vicinity of $\theta=0^{\circ}\pmod{180^{\circ}}$. One may also notice that $I_{S_1}^{~}$ still exhibits a maximum near $\theta=0$ at $E_{\rm L}=1.96~\rm eV$, but not at $E_{\rm L}=2.33~\rm eV$, where a maximum is found near $\theta=30^{\circ}$.

First, our observations provide useful empirical guidelines to determine the crystal orientation. In practice, the Re chains are most currently found to coincide with one of the cleaved edges of the $N$-layer ReX$_2$ crystals, providing a first hint of the crystal orientation. Then, at $E_{\rm L}=1.96 \rm ~eV$, $I_{S_1}^{~}$  reaches a maximum near $\theta=0^{\circ}$ both for ReSe$_2$ ad ReS$_2$. Nevertheless, we should stress that the present experimental results do not make it possible to unambiguously correlate the $S_{1}^{\pm}$ mode-features with the direction of the in-plane rigid layer displacements with respect to the Re chains. This observation invites further theoretical analysis of the normal displacements associated with the interlayer phonon modes.

Second, the diverse angular patterns shown in Fig.~\ref{Fig5}-\ref{Fig7}  and in the Supporting Information shed light on the extremely complex Raman response of ReX$_2$. In order to model the angular patterns described above, one has to consider the Raman tensor in ReX$_2$ (see Ref.~\citenum{Wolverson2014} and the Supporting Information). Since these materials efficiently absorb visible light~\cite{Ho1997,Ho1998,Ho2004,Zhong2015,Aslan2015}, the Raman tensor elements are complex numbers~\cite{Ribeiro2015,Kim2015}, which are dependent on the incoming laser photon energy. Using the general form for a symmetric, second rank tensor, we are able to fit the experimental angular dependence of the Raman intensities (see Fig.~\ref{Fig5}-\ref{Fig7} and Supporting Information). Moreover, our observation of distinct angular patterns on ReSe$_2$ and ReS$_2$ bilayers at two close-lying photon energies points towards resonance effects. We note that a recent theoretical study by Zhong \textit{et al.}  predicts a complex manifold of strongly anisotropic excitons in monolayer ReSe$_2$ and ReS$_2$, with several associated optical resonances in the spectral window (1.96~eV – 2.33~eV) investigated here~\cite{Zhong2015}. A further theoretical analysis of anisotropic exciton-phonon coupling together with an experimental study of the anisotropic dielectric function and resonance Raman profile in $N$-layer ReX$_2$ is mandatory to rationalize the anomalous photon energy dependence of the anisotropic Raman response unveiled in Fig.~\ref{Fig6} and \ref{Fig7}.

Third, we would like to emphasize that due to the photon energy dependence of the anisotropic Raman response in $N$-layer ReX$_2$, great care must be taken while correlating the angular dependence of the Raman response of a $N$-layer ReX$_2$ sample to its crystal orientation. In particular, the behavior of $I_{\rm V}^{~}$ shown in Fig.~\ref{Fig6}(c) restricts the applicability of the empirical analysis of Chenet \textit{et al.}~\cite{Chenet2015}, which remains valid for ReS$_2$ at $E_{\rm L}=1.96~\rm eV$ and $E_{\rm L}=2.33~\rm eV$ and for ReSe$_2$ at $E_{\rm L}=1.96~\rm eV$ but not at $E_{\rm L}=2.33~\rm eV$.

\section*{Conclusion}

We have reported the observation of prominent, interlayer Raman modes in distorted octahedral (1T') ReSe$_2$ and ReS$_2$. The splitting of the interlayer shear modes and the strong angular dependence of the Raman intensity of the interlayer shear and breathing modes are a direct fingerprint of the pronounced in-plane anisotropy in these materials. Noteworthy, the observation of similar intralayer Raman spectra and fan diagrams of the interlayer modes in $N$-layer ReSe$_2$ and ReS$_2$, suggests that the two materials studied here are (quasi-)isostructural, bringing a key contribution to a longstanding debate summarized in Ref.~\citenum{Feng2015}. Our findings will be useful to improve our understanding of other anisotropic layered materials such as black phosphorus~\cite{Ribeiro2015,Ling15,Wu2015,Luo2015,Kim2015}. Finally, $N$-layer ReSe$_2$ and ReS$_2$ appear as a stimulating playground for further investigations of symmetry-dependent resonant exciton-phonon coupling~\cite{Carvalho2015,Scheuschner2015,Lee2015,Froehlicher2015}, whereas from a more applied perspective the anisotropic optical response~\cite{Ho2004,Aslan2015,Zhong2015} and  electron transport properties~\cite{Tiong1999,Liu2015,Lin2015} of these systems hold promise for  novel optoelectronic devices.

\textit{Note: while finalizing this manuscript and during the review process, we became aware of related Raman studies on ReS$_2$~\cite{Nagler2015,He2016,Qiao2015}. Although no splitting of the interlayer shear mode is resolved in Ref.~\citenum{Nagler2015}, Ref.~\citenum{He2016} reports a shear mode splitting of $\approx 4\rm~cm^{-1}$ in bilayer ReS$_2$. The discrepancy between the $\lesssim 1\rm~cm^{-1}$ splittings observed here and Ref.~\citenum{He2016} is likely due to different stacking orders with nearly equal stability, \textit{i.e.}, polytypism~\cite{Qiao2015}. Here, the stacking order in our $N$-layer ReS$_2$ crystals would correspond to the type 3 stacking studied in Ref.~\citenum{He2016}. Noteworthy, this stacking displays an inversion center for any number of layers.}

\section*{Methods}

$N$-layer ReX$_2$ crystals were prepared by mechanical exfoliation of commercially available bulk crystals (2D semiconductors) onto Si wafers covered with a 90-nm or 285-nm-thick SiO$_2$ epilayer. Our ReX$_2$ samples remained stable for several months in ambient conditions. The number of layers was first estimated using optical contrast and atomic force microscopy, and unambiguously confirmed by the present Raman study. Micro-Raman scattering measurements were carried out in ambient conditions, in a backscattering geometry using a monochromator equipped with a 2400 grooves/mm holographic grating, coupled to a two-dimensional liquid nitrogen cooled charge-coupled device (CCD) array.  The samples were excited using a linearly polarized laser at two different photon energies ($E_{\rm L}=2.33~\rm eV$ and $E_{\rm L}=1.96~\rm eV$). These energies are significantly above than the band-edge direct excitons that are near 1.3~eV and 1.5~eV in bulk ReSe$_2$ and ReS$_2$, respectively~\cite{Ho1997,Ho1998,Ho2004,Aslan2015,Tongay2014}. A moderate laser intensity below $50~ \rm kW/cm^2$ was used for all measurements. Spectral resolutions of $0.6~\rm cm^{-1}$ and $0.4~\rm cm^{-1}$ were obtained at $E_{\rm L}=2.33~\rm eV$ and $E_{\rm L}=1.96~\rm eV$, respectively. In order to attain the low-frequency range, a combination of one narrow bandpass filter and two narrow notch filters (Optigrate) was used. After optimization, Raman features at frequencies as low as 4.5~cm$^{-1}$ could be measured. In order to uncover the anisotropic Raman response of ReX$_2$ layers, the samples were rotated with respect to the fixed linear polarization of the incoming laser field. In practice, each sample was mounted onto a rotating stage attached onto a XYZ piezo-scanner and placed as close as possible to the rotational axis. The region of interest was then parked under the fixed, diffraction limited laser spot using the piezo stage. After each rotation step, the position of region of interest was re-adjusted under the laser spot (with an accuracy better than 100 nm) using the piezo-stage. Polarization-dependent Raman studies were performed using an analyzer placed before the entrance of our spectrometer, followed by a polarization scrambler in order to avoid any artifacts due to polarization sensitive optics in our spectrometer. Thereafter, we will only consider parallel (XX)  polarizations of the incoming and Raman-scattered photons. Data recorded in the perpendicular (XY) configuration are shown in the Supporting Information. Finally, the measured Raman features are fit to Voigt profiles, taking into account our spectral resolution.


\begin{acknowledgements}
We thank F. Ben Romdhane and F. Banhart for recording transmission electron microscopy images of a bulk ReSe$_2$ crystal, F. Fernique for early contributions to this study, and the StNano clean room staff for experimental help. We acknowledge financial support from the Agence Nationale de la Recherche (under grant QuanDoGra 12 JS10-001-01 and H2DH), from the CNRS and Universit\'e de Strasbourg, and from the LabEx NIE (under grant WHO). 
\end{acknowledgements}


%


\newpage
\begin{center}
{\LARGE\textbf{Supplementary Information}}
\end{center}

\setcounter{equation}{0}
\setcounter{figure}{0}
\setcounter{section}{0}
\renewcommand{\theequation}{S\arabic{equation}}
\renewcommand{\thefigure}{S\arabic{figure}}
\renewcommand{\thesection}{S\arabic{section}}
\linespread{1.4}

This document contains the following sections:
\begin{itemize}
\item Determination of the crystal orientation in ReSe$_2$  
\item Additional Raman data on ReSe$_2$ 
\item Additional Raman data on ReS$_2$  
\item Angular dependence of the Raman intensities  
\end{itemize}

\clearpage


\section{Determination of the crystal orientation in \element{Re}\element{Se}$_2$}
\label{S1}

\begin{figure}[!htb]
\begin{center}
\includegraphics[width=1\linewidth]{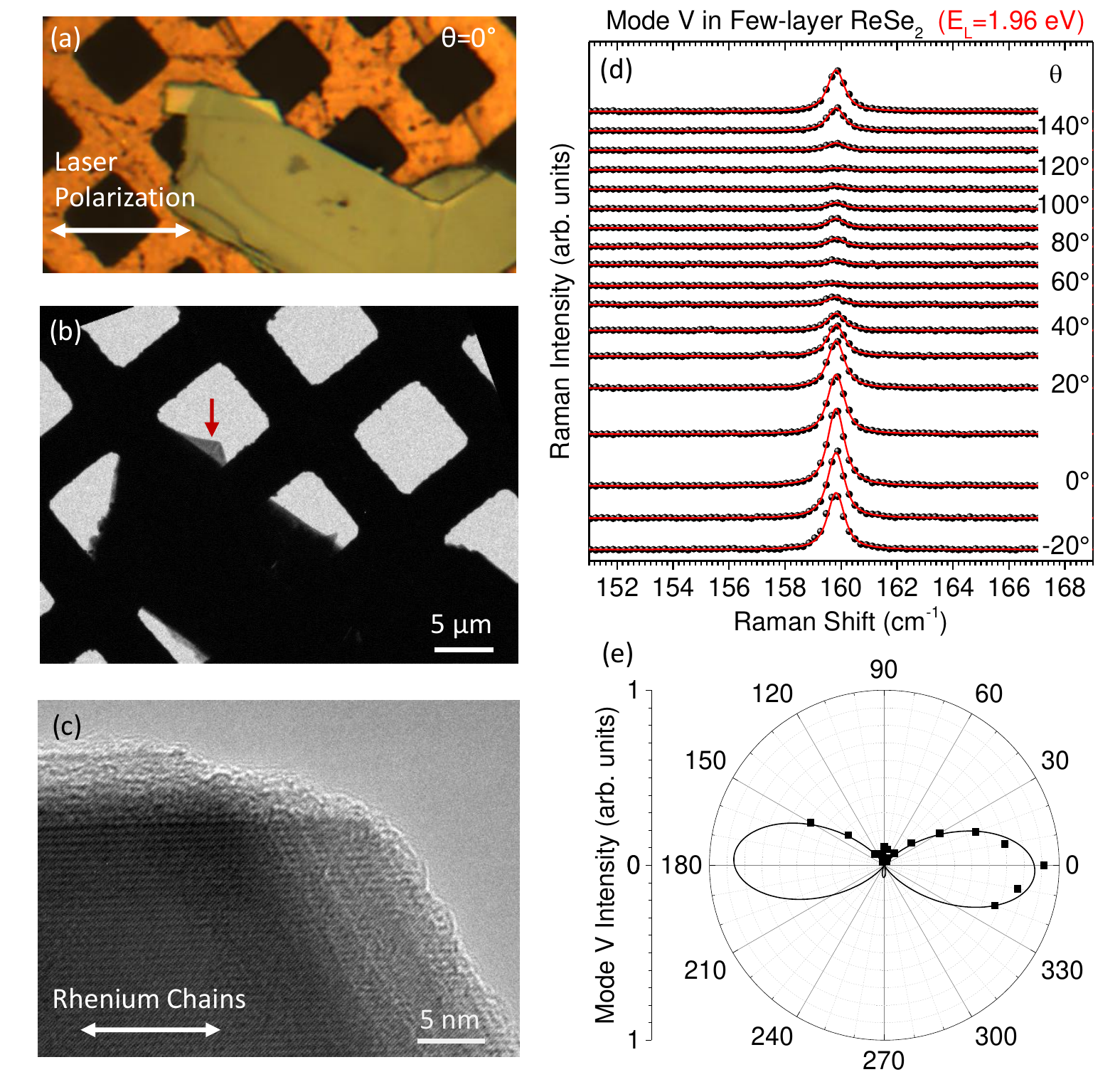}
\caption{(a) Optical and (b) low-resolution TEM image of a ReSe$_2$ crystal placed onto a TEM grid. (c) High-resolution TEM image of the region highlighted with the red arrow in (b), showing the Re Chains. (d) Angular dependence of the Raman mode V recorded as a function of the angle $\theta$ between the linearly polarized laser field and the Re chains on the region highlighted in (b) in the parallel (XX) polarization configuration. The raw spectra (spheres) are fit to Voigt profiles (red lines) and are vertically offset for clarity. (e) Polar plot of the Raman mode V integrated intensity. The solid lines in (e) is a fit based on Eq.~\eqref{eqXX}. $\theta=0$ corresponds to a laser polarization parallel to the Re chains.}
\label{FigS1}
\end{center}
\end{figure}

To verify whether the empirical approach introduced by Chenet \textit{et al.} on ReS$_2$~\cite{Chenet2015} is still holding for ReSe$_2$ we recorded the angular dependence of the intensity of the mode V at $168~\rm cm^{-1}$ in a few-layer ReSe$_2$ crystal. Prior to the Raman measurements, the ReSe$_2$ crystal was transferred onto a transmission electron microscopy (TEM) grid (see Fig.~\ref{FigS1}(a), with one freestanding region that could be imaged using by TEM (see Fig.~\ref{FigS1}(b),(c)), allowing a direct observation of the Re chains. Raman measurements were then performed on the same region at $E_{L}=1.96~\rm eV$. As shown in Fig.\ref{FigS1}(d),(e), the integrated intensity of the mode V is indeed maximal when the laser polarization is parallel to the Re chains. This empirical approach was then used, only at $E_{L}=1.96~\rm eV$, to confirm the orientation of the Re Chains in the $N$-layer ReSe$_2$ samples shown in the main manuscript.

For ReS$_2$ samples, we relied on the results by Chenet \textit{et al.}, obtained at $E_{L}=2.33~\rm eV$ and found that similar angular dependencies of the mode V integrated intensities were obtained at $E_{L}=2.33~\rm eV$ and at $E_{L}=1.96~\rm eV$.

\clearpage

\section{Additional Raman data on \element{Re}\element{Se}$_2$}
\label{S2}

\begin{figure}[!htb]
\begin{center}
\includegraphics[width=1\linewidth]{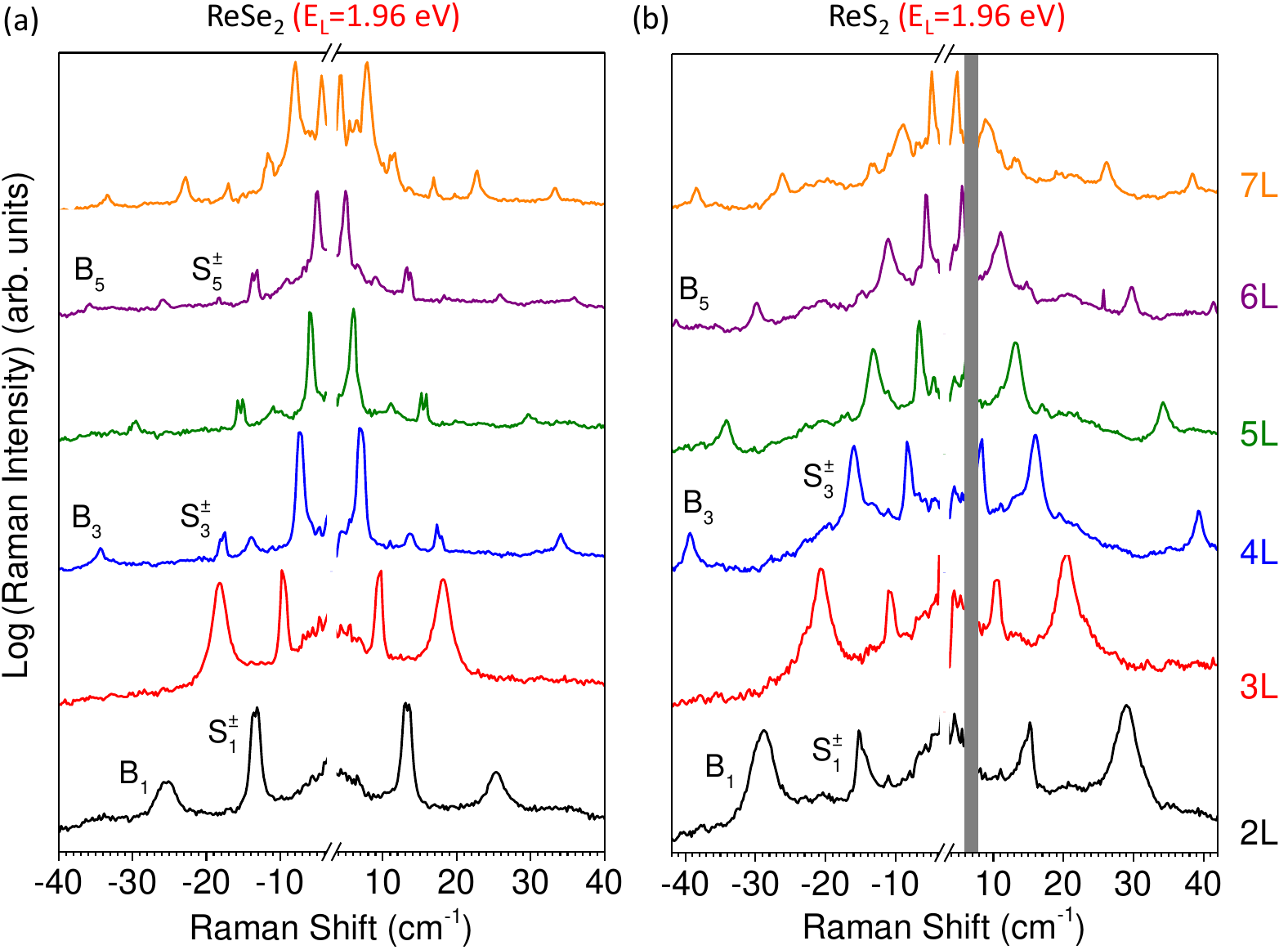}
\caption{Ultralow frequency Raman spectra of $N=2$ to $N=7$ layers (a) ReSe$2$ and (b) ReS$_2$, recorded at $E_{\rm L}=1.96~\rm eV$ in the XX configuration, showing the LSM and the LBM features. The grey bar in (b) masks a residual contribution from the laser beam. The mode labels correspond to the notations in Fig.~3 in the main manuscript.}
\label{FigS2}
\end{center}
\end{figure}

\begin{figure}[!htb]
\begin{center}
\includegraphics[width=1\linewidth]{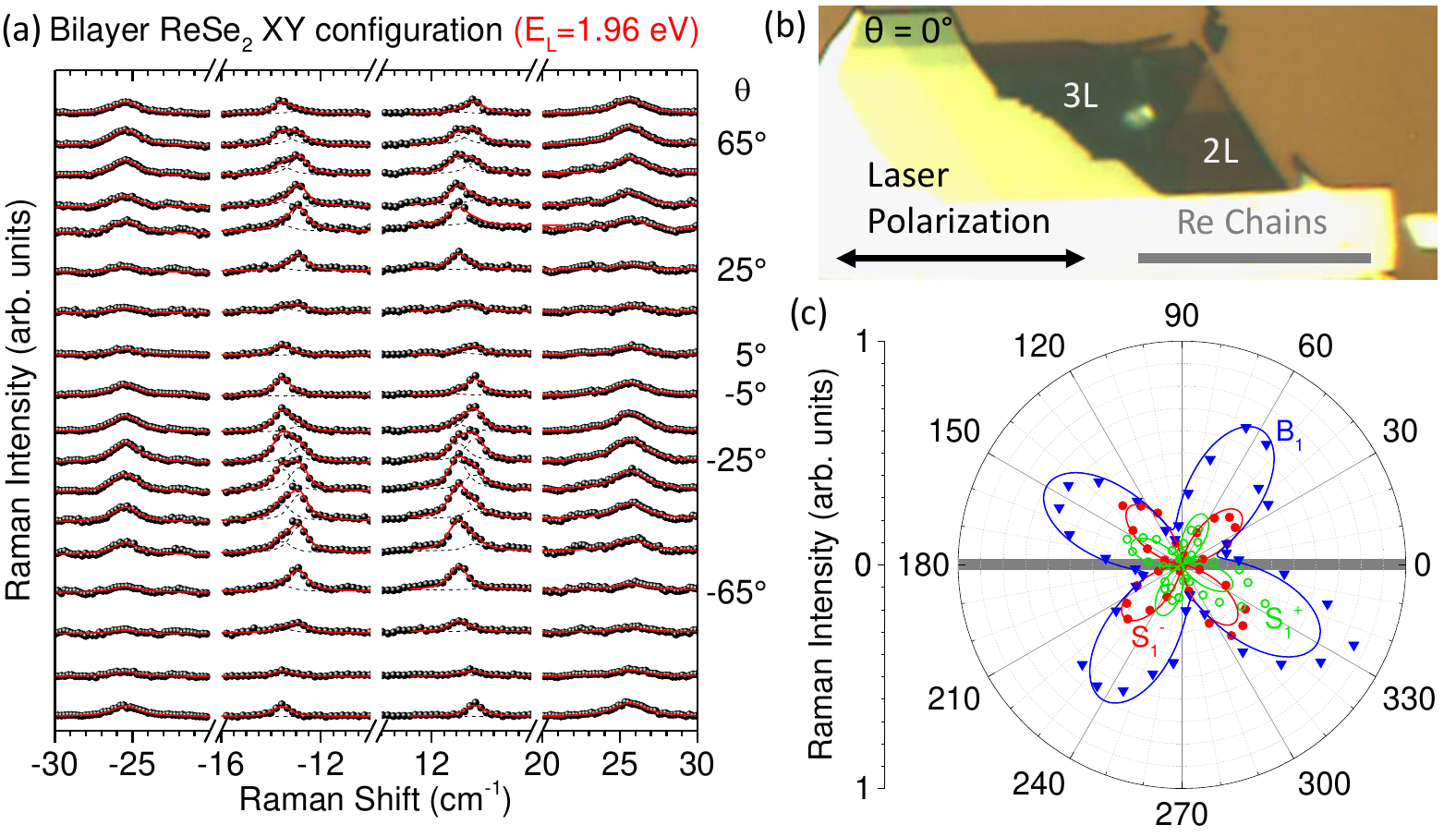}
\caption{(a) Angular dependence of the low-frequency Raman spectra of a ReSe$_2$ bilayer recorded as a function of the angle $\theta$ between the linearly polarized laser field and the Re chains at $E_{\rm L}=1.96~\rm eV$ in the perpendicular (XY) configuration.  The raw spectra (spheres) are fit to Voigt profiles (red lines) and are vertically offset for clarity. (b) Optical image of the ReSe$_2$ bilayer showing the direction of the rhenium chains. At $\theta=0$, the laser polarization  is parallel to the Re chains. (c) Polar plot of the integrated Raman intensity of the $S_1^{\pm}$ and $B_1^{~}$ modes. The solid lines in (c) are fits to the experimental data based on Eq.~\eqref{eqXY}.}
\label{FigS3}
\end{center}
\end{figure}

\begin{figure}[!htb]
\begin{center}
\includegraphics[width=1\linewidth]{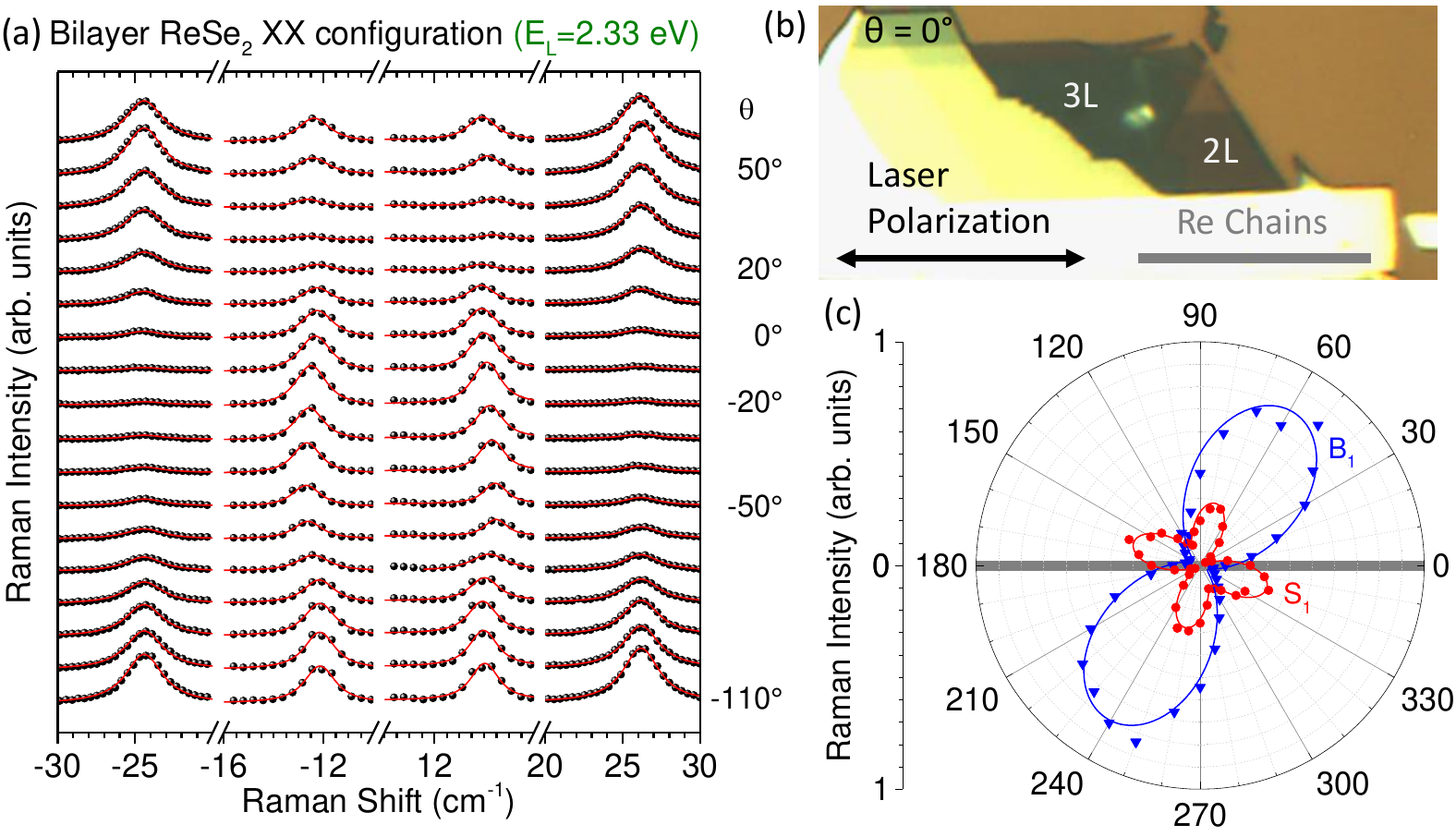}
\caption{(a) Angular dependence of the low-frequency Raman spectra of a ReSe$_2$ bilayer recorded as a function of the angle $\theta$ between the linearly polarized laser field and the Re chains at $E_{\rm L}=2.33~\rm eV$ in the parallel (XX) configuration.  The raw spectra (spheres) are fit to Voigt profiles (red lines) and are vertically offset for clarity. (b) Optical image of the ReSe$_2$ bilayer showing the direction of the rhenium chains. The laser polarization for the reference at $\theta=0$ is parallel to the Re chains. (c) Polar plot of the integrated Raman intensity of the $S_1^{\pm}$ and $B_1^{~}$ modes. The solid lines in (c) are fits to the experimental data based on Eq.~\eqref{eqXX}.}
\label{FigS4}
\end{center}
\end{figure}

\begin{figure}[!htb]
\begin{center}
\includegraphics[width=1\linewidth]{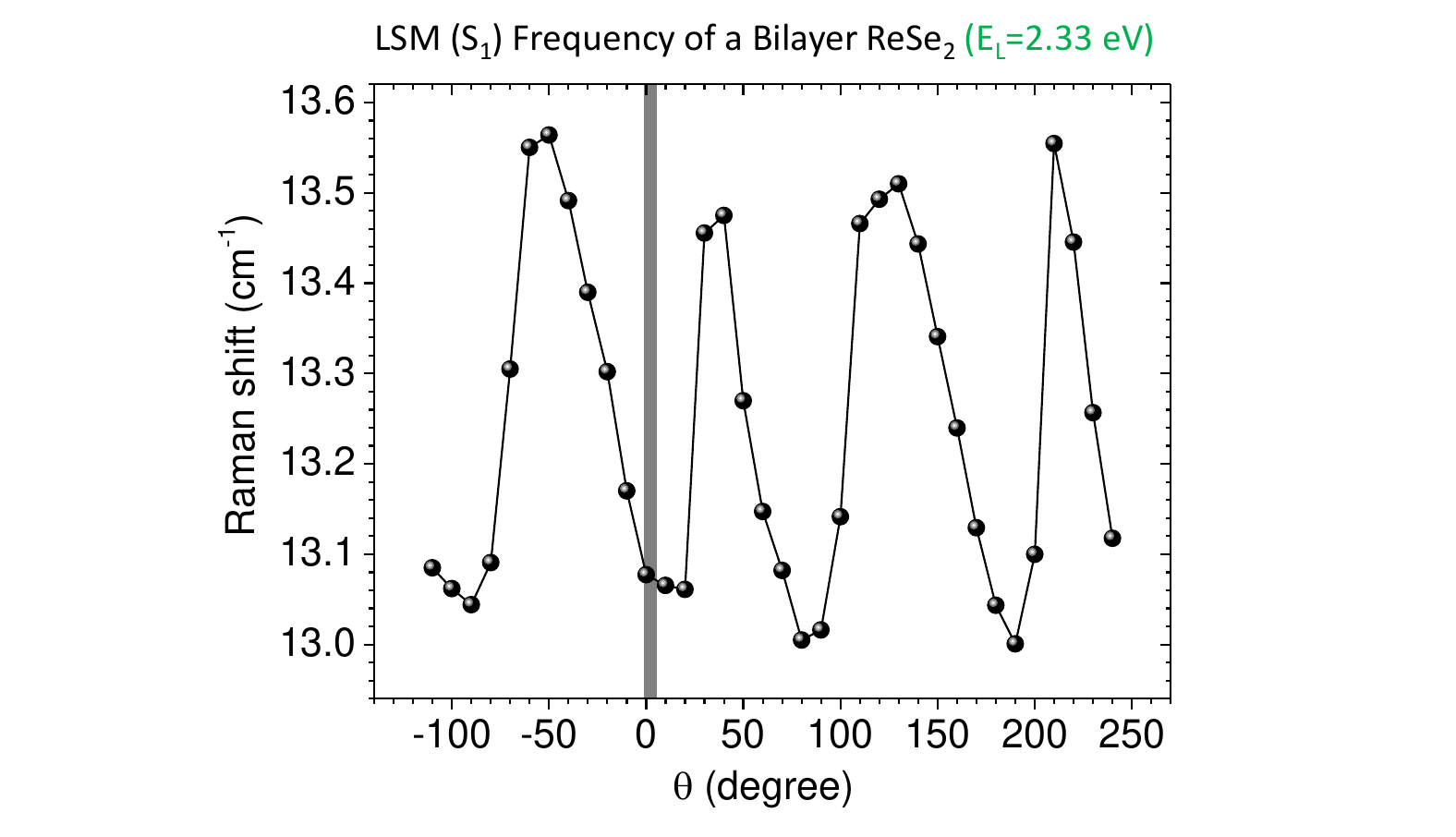}
\caption{Angular dependence of the frequency of the LSM in a ReSe$_2$ bilayer recorded as a function of the angle $\theta$ between the linearly polarized laser field and the Re chains at $E_{\rm L}=2.33~\rm eV$ in the parallel (XX) configuration. At $\theta=0$, the laser polarization  is parallel to the Re chains. The frequency of the LSM is minimal when the laser polarization is parallel or perpendicular to the Re chains and reaches a maximum when the laser polarization makes an angle $\theta=45^{\circ}\pmod{90^{\circ}}$ with the Re chains. This observation provides a useful criterion to identify the Re chains.}
\label{FigS6}
\end{center}
\end{figure}

\begin{figure}[!htb]
\begin{center}
\includegraphics[width=1\linewidth]{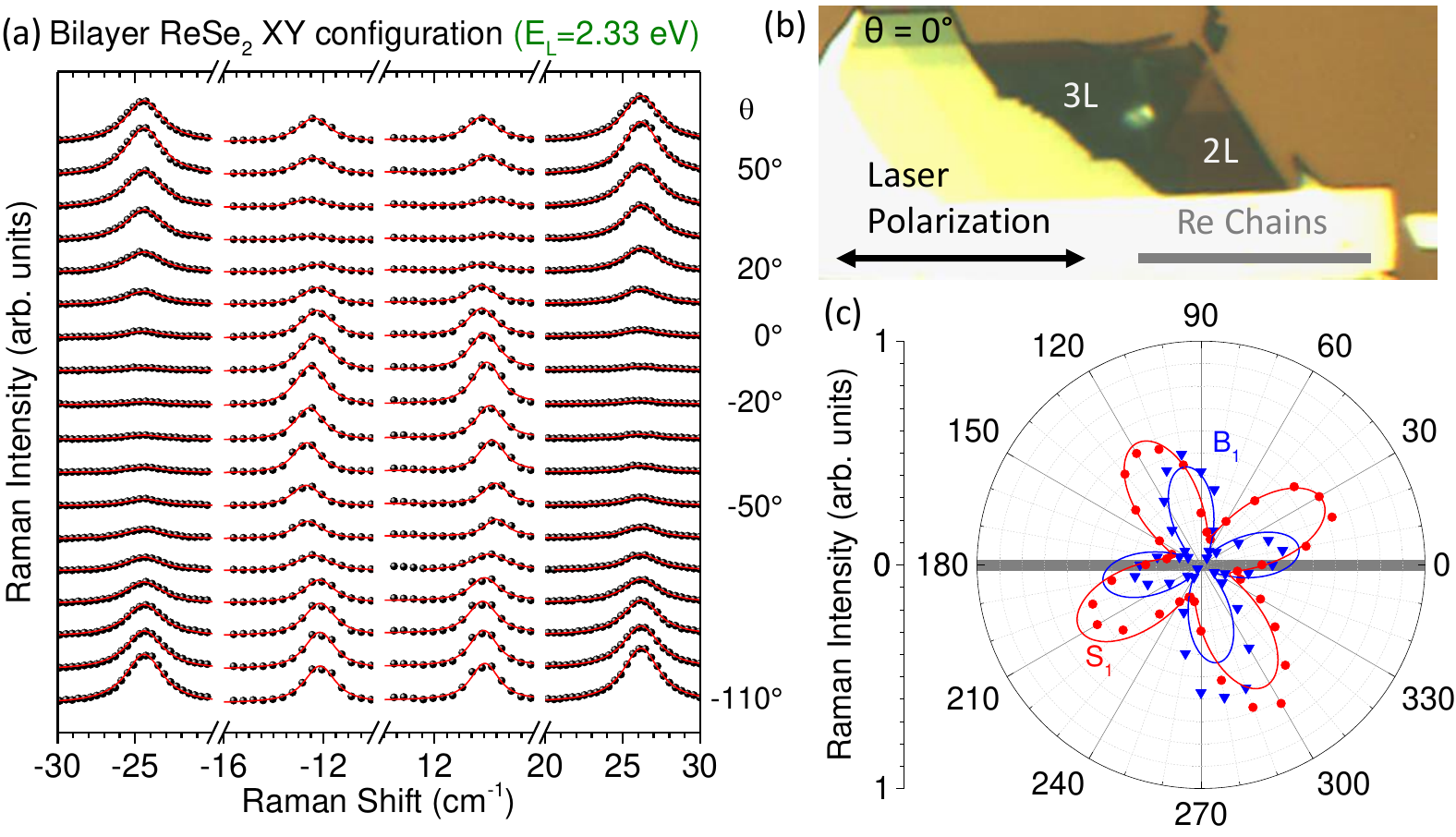}
\caption{(a) Angular dependence of the low-frequency Raman spectra of a ReSe$_2$ bilayer recorded as a function of the angle $\theta$ between the linearly polarized laser field and the Re chains at $E_{\rm L}=2.33~\rm eV$ in the perpendicular (XY) configuration.  The raw spectra (spheres) are fit to Voigt profiles (red lines) and are vertically offset for clarity. (b) Optical image of the ReSe$_2$ bilayer showing the direction of the rhenium chains. At $\theta=0$, the laser polarization  is parallel to the Re chains. (c) Polar plot of the integrated Raman intensity of the $S_1^{\pm}$ and $B_1^{~}$ modes. The solid lines in (c) are fits to the experimental data based on Eq.~\eqref{eqXY}.}
\label{FigS5}
\end{center}
\end{figure}

\begin{figure}[!htb]
\begin{center}
\includegraphics[width=1\linewidth]{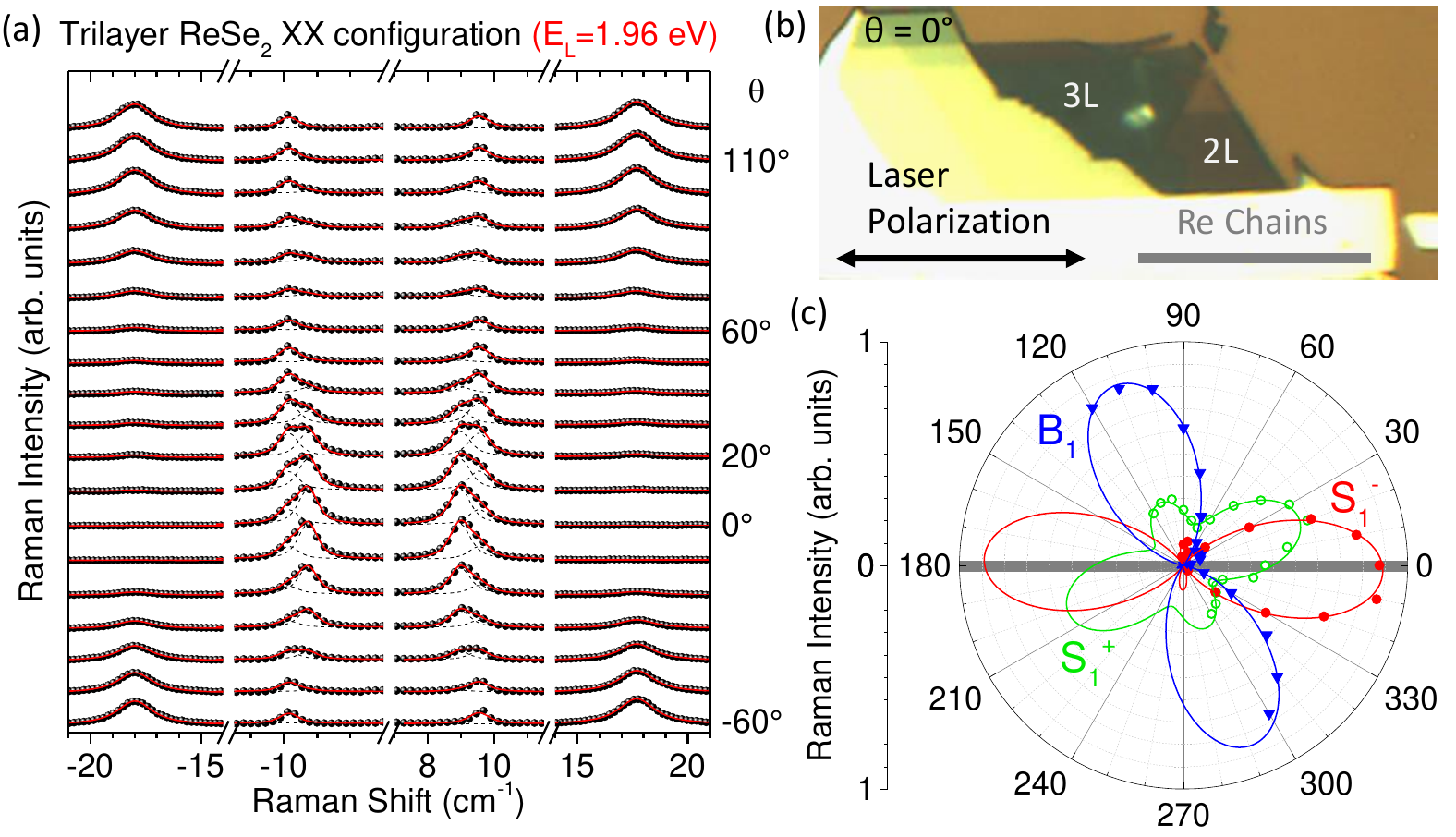}
\caption{(a) Angular dependence of the low-frequency Raman spectra of a ReSe$_2$ trilayer recorded as a function of the angle $\theta$ between the linearly polarized laser field and the Re chains at $E_{\rm L}=1.96~\rm eV$ in the parallel (XX) configuration.  The raw spectra (spheres) are fit to Voigt profiles (red lines) and are vertically offset for clarity. (b) Optical image of the  ReSe$_2$ trilayer showing the direction of the rhenium chains.  At $\theta=0$, the laser polarization is parallel to the Re chains. (c) Polar plot of the integrated Raman intensity of the $S_1^{\pm}$ and $B_1^{~}$ modes. The solid lines in (c) are fits to the experimental data based on Eq.~\eqref{eqXX}.}
\label{FigS7}
\end{center}
\end{figure}


\clearpage

\section{Additional Raman data on \element{Re}\element{S}$_2$}
\label{S3}

\begin{figure}[!htb]
\begin{center}
\includegraphics[width=1\linewidth]{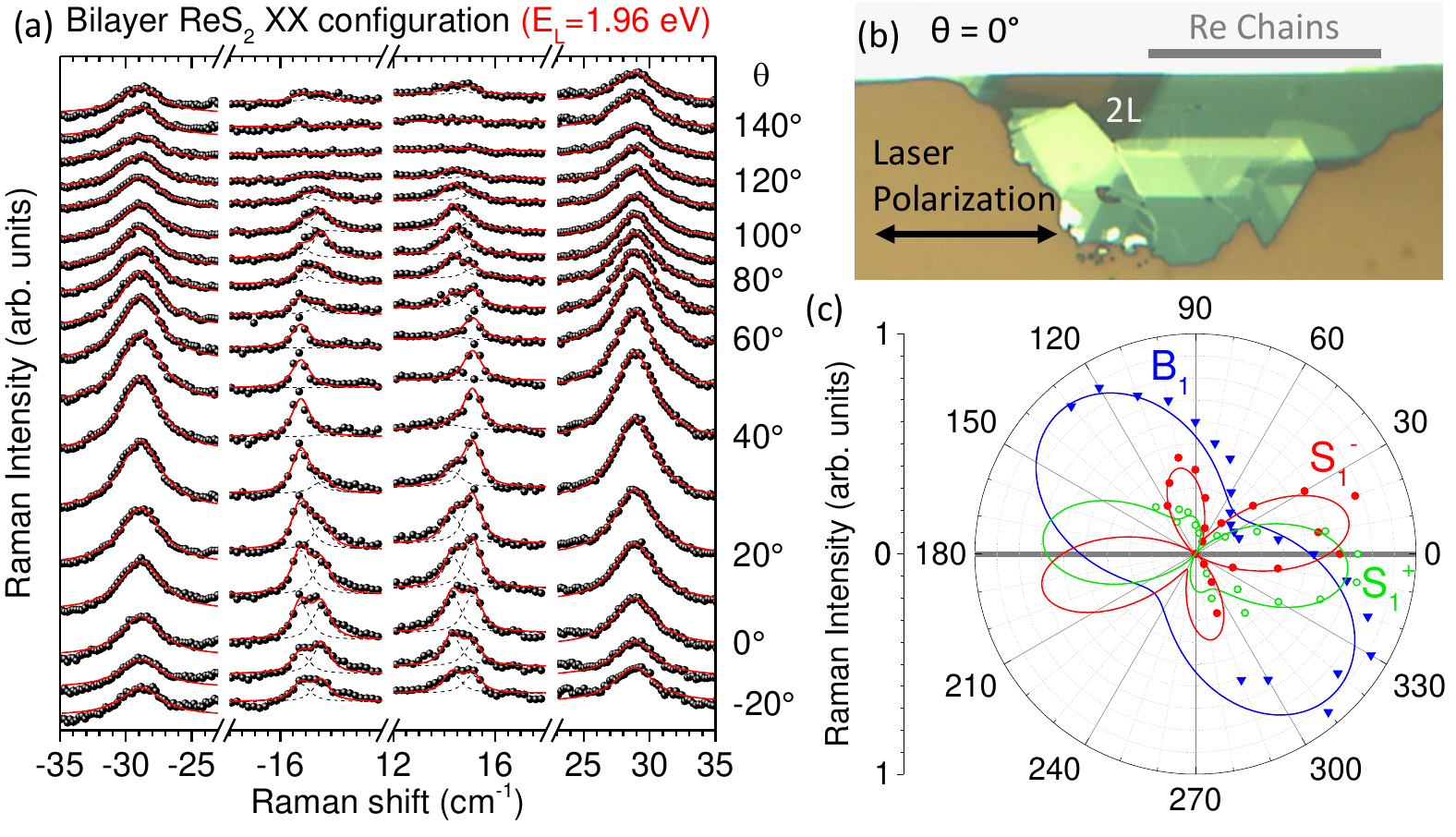}
\caption{(a) Angular dependence of the low-frequency Raman spectra of a ReS$_2$ bilayer recorded as a function of the angle $\theta$ between the linearly polarized laser field and the Re chains at $E_{\rm L}=1.96~\rm eV$ in the parallel (XX) configuration.  The raw spectra (spheres) are fit to Voigt profiles (red lines) and are vertically offset for clarity. (b) Optical image of the ReS$_2$ bilayer showing the direction of the rhenium chains.  At $\theta=0$, the laser polarization  is parallel to the Re chains. (c) Polar plot of the integrated Raman intensity of the $S_1^{\pm}$ and $B_1^{~}$ modes. The solid lines in (c) are fits to the experimental data based on Eq.~\eqref{eqXX}.}
\label{FigS9}
\end{center}
\end{figure}

\begin{figure}[!htb]
\begin{center}
\includegraphics[width=1\linewidth]{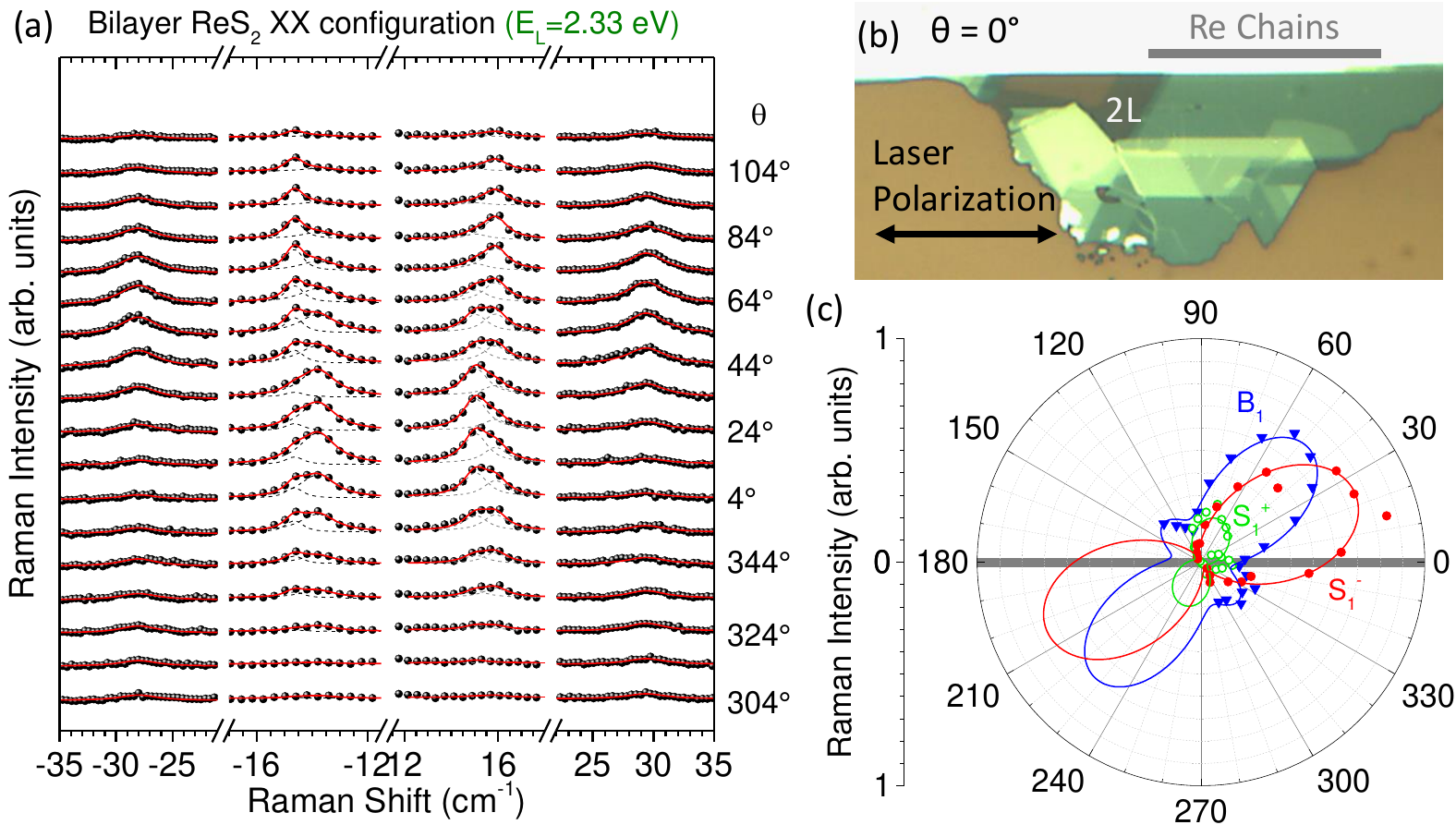}
\caption{(a) Angular dependence of the low-frequency Raman spectra of a ReS$_2$ bilayer recorded as a function of the angle $\theta$ between the linearly polarized laser field and the Re chains at $E_{\rm L}=2.33~\rm eV$ in the parallel (XX) configuration.  The raw spectra (spheres) are fit to Voigt profiles (red lines) and are vertically offset for clarity. (b) Optical image of the ReS$_2$ bilayer showing the direction of the rhenium chains. At $\theta=0$, the laser polarization  is parallel to the Re chains. (c) Polar plot of the integrated Raman intensity of the $S_1^{\pm}$ and $B_1^{~}$ modes. The solid lines in (c) are fits to the experimental data based on Eq.~\eqref{eqXX}.}
\label{FigS10}
\end{center}
\end{figure}

\clearpage
\section{Angular dependence of the Raman intensities}
\label{S4}

Since the optical absorption spectrum of ReX$_2$ is dominated by strongly anisotropic excitons~\cite{Aslan2015,Zhong2015}, the Raman response of these systems is expected to exhibit a complex angular dependence~\cite{Wolverson2014}. For an incoming and scattered polarizations (in the XY plane) denoted by ${\bm E}_i$ and ${\bm E}_s$, the Raman intensity writes

\begin{equation}
I(\theta,\alpha) \propto \left| {\bm E}_i \cdot \mathcal{ \bm R} \cdot {\bm E}_s \right|^2
\end{equation}
with $\theta$, the angle between ${\bm E}_i$ and the rhenium chains, $\alpha$ the angle between ${\bm E}_s$ and ${\bm E}_i$  and the Raman tensor 
\begin{equation}
\mathcal{R}=\begin{pmatrix}
   u\:e^{i\phi_u} & v\:e^{i\phi_v} \\
   v\:e^{i\phi_v} & w\:e^{i\phi_w} 
\end{pmatrix}
\end{equation}
with $u$ ($\phi_u$), $v$ ($\phi_v$), $w$ ($\phi_w$) the amplitude and phase of the Raman tensor elements. As introduced in Ref.~\cite{Ribeiro2015}, the finite phases stem the anisotropic absorption of ReX$_2$.

In the parallel configuration ($\alpha=0$), the Raman intensity writes

\begin{multline}
I_{\rm XX}(\theta) \propto u^2 \:\cos^4\theta + w^2\: \sin^4\theta  + 4v^2\: \cos^2\theta\:\sin^2\theta \\
+2\:uw\:\cos\phi_{uw}\: \cos^2\theta\:\sin^2\theta\\
+4\:uv\:\cos\phi_{uv}\: \cos^3\theta\:\sin\theta+ 4\:vw\:\cos\phi_{vw}\: \cos\theta\:\sin^3\theta 
\label{eqXX}
\end{multline}
with $\phi_{uv}=\phi_u-\phi_v$, $\phi_{uw}=\phi_u-\phi_w$, and $\phi_{vw}=\phi_v-\phi_w$.

In the perpendicular configuration ($\alpha=\pi/2$), the Raman intensity writes
\begin{multline}
I_{\rm XY}(\theta) \propto (u^2+w^2) \: \cos^2\theta\:\sin^2\theta +v^2\:(\sin^2\theta-\cos^2\theta)^2\\
-2\:uw\:\cos\phi_{uw}\:\cos^2\theta\:\sin^2\theta\\
+2\:(uv\:\cos\phi_{uv}-\:vw\:\cos\phi_{vw}) \: (\sin^3\theta\:\cos\theta-\cos^3\theta\:\sin\theta)
\label{eqXY}
\end{multline}

Eq.~\eqref{eqXX} and Eq.~\eqref{eqXY} lead to a simpler expression for the total Raman intensity $I_{\rm tot}=I_{\rm XX} + I_{\rm XY}$:
\begin{equation}
I_{\rm tot}(\theta)\propto u^2 \:\cos^2\theta + w^2\: \sin^2\theta  + v^2 + 2\:(uv\:\cos\phi_{uv}+vw\:\cos\phi_{vw})\: \cos\theta\:\sin\theta
\label{eqtot}
\end{equation}

In our experiments, we cannot fit all the experimentally observed angular patters in Fig.~5-7 using real tensor elements. In practice \eqref{eqXX} and \eqref{eqXY} are used to fit the experimental angular dependences of the Raman intensities. The observation of a very distinct angular dependence of the Raman intensities for two different laser photon energies  is a direct proof that the Raman tensor elements strongly depend on the laser photon energy.

\end{document}